\documentclass{ws-procs9x6}

\begin{document}

\title{EXPLORING UNIVERSALITY OF FEW-BODY PHYSICS BASED ON ULTRACOLD ATOMS NEAR FESHBACH RESONANCES}

\author{Nathan Gemelke, Chen-Lung Hung, Xibo Zhang and Cheng Chin$^*$}

\address{%
  James Franck Institute and Physics Department, University of Chicago,\\
  Chicago, IL 60637, USA\\
  $^*$E-mail: cchin@uchicago.edu\\
  Website: http://ultracold.uchicago.edu
}



\begin{abstract}

  A universal characterization of interactions in few- and many-body
  quantum systems is often possible without detailed description of
  the interaction potential, and has become a defacto assumption
  for cold atom research. Universality in this context is defined
  as the validity to fully characterize the system in terms of
  two-body scattering length. We discuss universality in the following
  three contexts: closed-channel dominated Feshbach resonance,
  Efimov physics near Feshbach resonances, and corrections to the
  mean field energy of Bose-Einstein condensates
  with large scattering lengths. Novel experimental tools and
  strategies are discussed to study universality in
  ultracold atomic gases: dynamic control of interactions,
  run-away evaporative cooling in optical traps, and preparation of
  few-body systems in optical lattices.

\end{abstract}


\bodymatter

\section{Introduction}

Quantum gases of ultracold atoms distinguish themselves from other
quantum systems in two unique and useful ways. First of all, the
diluteness of the gases permits a very simple and accurate
description of the effect of interactions. Degenerate gases of atoms
can be described well by textbook models of fundamental and general
interest. Extending beyond these, complexity can be built in slowly
to
study novel quantum phases, and even intractable mathematical
models. This aspect has inspired new research in the vein of
\emph{quantum simulation}, promising far-reaching impact on the
understanding of other quantum systems in nature, including
condensed and nuclear matter.

A second gainful aspect of ultracold atomic gases lies in the
ability to tune interactions via Feshbach resonant scattering.
Exploitation of this feature, first observed in 1998
\cite{Inouye1998}, has only fully matured in recent years, and
promises numerous future applications. Full control of interaction
in a quantum gas not only allows for an easy exploration of the
quantum system in different interaction regimes, but also leads to
new methods to observe dynamic evolution and to scrutinize
quantum states in previously unimaginable ways. For example,
projecting a complex many-body state onto a non-interacting single
particle basis can be realized by diabatically switching off atomic
interactions.

The majority of quantum gas systems studied to date admit a
universal description of the effect of interaction. In this paper,
we describe our approach to explore situations in which universality
requires nontrivial extensions. Our experimental platform, based on
optically trapped cesium atoms, exploits both of the aforementioned
features, allowing us to address long-standing questions concerning
the universality of an interacting gas. In particular, we will focus
on three topics: universality and its minimal extensions in the
study of dimer molecules, three-body Efimov states near Feshbach
resonance, and beyond mean-field interactions in Bose-Einstein
condensates. Finally, we will outline our approach to study few-body
interactions and our experimental progress.

\section{Universality in $N$-body physics}

The connection of quantum degenerate atomic gases to other physical
systems is made possible by the expected universality of physics at
low temperatures. Here, universality arises when the quantum system
is fully described by a single parameter, the two-body scattering
length $a$ \cite{Braaten2006}. Universality is well established in
two-body, low energy scattering theory, where the $s$-wave
scattering phase shift is $\eta=-\tan^{-1} ka$, with $k$ the
scattering wave number. In the many-body regime, universal behavior
of dilute Bose-Einstein condensates (BECs) of different bosonic
atomic species is expected for small and positive scattering
lengths. Universality is further expected and verified in
two-component degenerate Fermi gases with large scattering lengths,
as in the BEC-BCS (Bardeen-Cooper-Schrieffer superfluid) crossover
regime \cite{Regal2004,Bartenstein2004,Zwierlein2004}.

Non-universal parameters, however, can play an important role in
certain low energy few- and many-body systems, and represent the
entrance of a richer underlying scattering physics. For example,
binding energies of Efimov states in three-body systems
\cite{Efimov1970} and three-body interactions in BECs with large
scattering lengths \cite{Braaten2002} are expected to be
non-universal. Both cases strongly depend on the three-body
scattering phase shifts, which likely cannot be universally
derived from $a$ \cite{Braaten2006}.



\section{Feshbach resonances}
\subsection{Origin of Feshbach resonance}
In cold atom experiments, Feshbach resonances occur when two free
atoms interact in the scattering channel and resonantly couple to a
bound molecular state in a closed channel \cite{Tiesinga1993}. In
many cases, the bound state can have a different magnetic moment
than that of the scattering atoms, and resonant coupling between the
channels can be induced by tuning the bound state energy with an
external magnetic field.

Near a Feshbach resonance, the scattering phase shift $\eta$ follows
the Breit-Wigner formula \cite{Mott1965}:

\begin{equation}
 \eta=\eta_{\mathrm{bg}}-\tan^{-1}\frac{\Gamma/2}{E-E_c-\delta E} \,,
 \label{phaseshift}
\end{equation}

\noindent where $\eta_{\mathrm{bg}}$ is the background, or
off-resonant, phase shift, $E=\hbar^2k^2/m$ is the scattering
energy, $k$ is the scattering wave number, $m$ is twice the reduced
mass, $\Gamma\propto k$ is the coupling strength between the
scattering and bound states, $E_c$ is the energy of the bare bound
state, and $\delta E$ is the self-energy shift.

At low scattering energies $E\rightarrow 0$, the (background)
scattering length is given by $a_{(\mathrm{bg})}=-\tan
\eta_{(\mathrm{bg})}/k$ \cite{Mott1965}. We further assume a linear
Zeeman shift to the bound state $E_c=\delta\mu(B-B_c)$, where
$\delta\mu$ is the relative magnetic moment between open and closed
channels, and the bound state is shifted to the continuum when
$B=B_c$. These allow us to derive scattering length in the standard
resonance form $a=a_{\mathrm{bg}}[1-\Delta/(B-B_0)]$. Here
$\Delta=\lim_{k\rightarrow 0}\Gamma/(2ka_{\mathrm{bg}}\delta\mu)$ is
the resonance width, $B_0=B_c-\delta E/\delta\mu$ is the resonance
position. Note that $a$ diverges when $B=B_0$, or equivalently,
$\eta=(N+\frac12)\pi$, where $N$ is the number of molecular states
below the continuum.

It is important to note that Feshbach resonance occurs not exactly
when the bare state is tuned to the continuum $B=B_c$. From the van
der Waals potential model, the resonance position offset is
\cite{Julienne2006}

\begin{equation}
 B_0-B_c =-\frac{\delta E}{\delta\mu}=-\frac{r^2-r}{r^2-2r+2}\Delta
 \, ,
  \label{b0}
\end{equation}

\noindent where $r=a_{\mathrm{bg}}/\bar{a}$ and $\bar{a}$ is the
mean scattering length of the van der Waals potential
\cite{Gribakin1993}.

Eq.~\ref{b0} shows that the difference in magnetic field between the
bare state crossing and the resonance position $B_0-B_c$ is on the
order of the resonance width $\Delta$ when $|a_{\mathrm{bg}}|>
\bar{a}$.

\subsection{Non-universality of Feshbach molecules}

\label{f} 
\begin{figure}
\begin{center}
\psfig{file=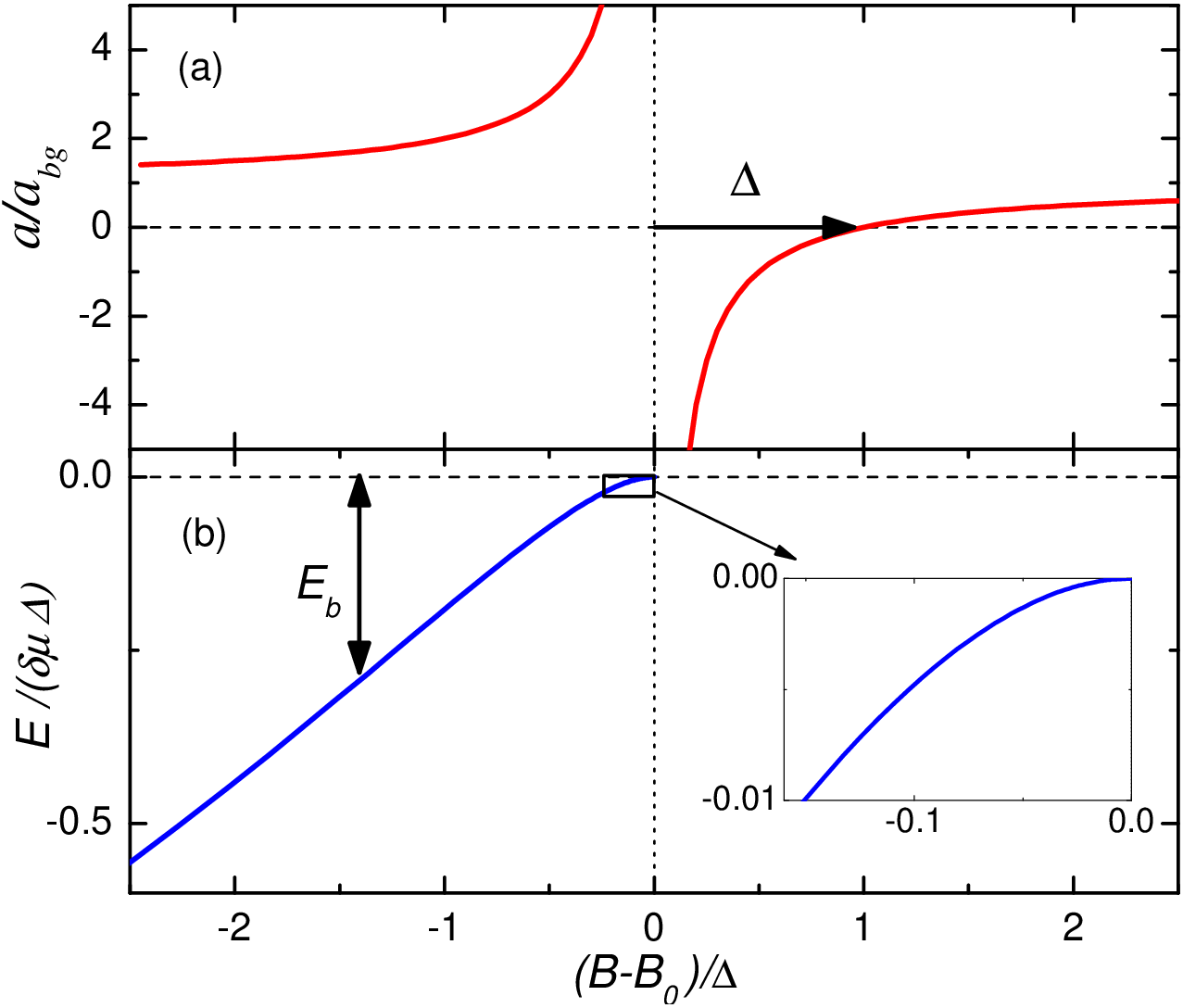,width=2.5in} \caption{Scattering length $a$
in (a) and molecular state energy $E$ in (b) near a magnetically
tuned Feshbach resonance. $E_b>0$ is the binding energy.  The inset
shows the universal regime where $E_b=\hbar^2/m a^2$.} \label{e&a}
\end{center}
\end{figure}

The two-channel nature of the interaction potential
described in the previous section implies that, in the absence of
Feshbach coupling, the scattering length of atoms in the entrance
channel does not reveal the properties of the closed channel bound
state. Thus, the properties of the molecular state are clearly
non-universal. This point can also been seen in Fig.~\ref{e&a}. When
the molecular state is well below the continuum, the molecular
energy approaches the bare state value $E_c=\delta\mu(B-B_c)$, which
cannot be universally derived from mere knowledge of $a$.

When the magnetic field is tuned sufficiently near the Feshbach
resonance, Feshbach coupling strongly modifies the nature of the
bound state, whose character is now dominated by the open channel.
In this regime, the molecular state does develop a universal
behavior with a binding energy of $E_b=\hbar^2/m a^2$. The Universal
regime can be seen in Fig.~\ref{e&a} (b) inset.

The transition between the non-universal and universal regimes
differs for different resonances. One instructive way to see the
transition behavior is to expand the molecular binding energy near
the resonance, which, from a simplified two-channel model potential
\cite{Chin2008}, gives

\begin{equation}
 E_b\approx\frac{\hbar^2}{m(a-\bar{a}-R^*)^2} \, .
  \label{eb}
\end{equation}

\noindent Here $\bar{a}$, $R^*=a_{\rm bg}E_{\rm
bg}/(\delta\mu\Delta)$ and $E_{\rm bg}=\hbar^2/ma_{\rm bg}^2$
\cite{Petrov2004}, are two leading order non-universal length scales
associated with the finite interaction range and the coupling
strength to the closed channel bound state, respectively.
Universality is valid only when $a\gg \bar{a}$ and $R^*$. In
particular, $R^*$, can be uncharacteristically large, $R^*\gg
\bar{a}$, for narrow resonances . Resonances of this type are deemed
closed-channel dominated and can have strong non-universal behavior.

\section{Efimov physics and Efimov states}
Efimov states are a set of three-body, long-range bound states which
emerge when the pairwise interactions in a three-particle system are
resonantly enhanced. These states are supported by the Efimov
potential which scales like $-1/R^2$ for $R<|a|$, where the
hyperspherical radius $R$ characterizes the geometric size of the
system \cite{Efimov1970}.

The connection between the three-body Efimov potential and
scattering length $a$ can be understood using a hand-waving picture.
Assume two bosonic atoms are separated by $R$, the wave function of
the third atom is scattered by each atom, with $|a|$ characterizing
the length scale of the scattered waves. The total wave function
$\psi$, after Bose-symmetrization, can be significantly enhanced
when the two scattered waves overlap. Using Schr\"{o}dinger's
equation, we can model the wave function enhancement
$\phi(R)=\delta\psi(R)$ as a result of an effective Efimov potential
$V_{\mathrm{efm}}(R)$, which satisfies Schrodinger's equation:

\begin{equation}
-\frac{\hbar^2}m\phi''(R)+V_{\mathrm{efm}}(R)\phi(R)=E\phi(R)\,\,\,\,\,\,\mbox{for
$R<a$}.
  \label{s_eq}
\end{equation}

To evaluate the curvature term, we note that $\phi$ is localized
with a length scale of system size $R$. The curvature is thus
negative and we can rewrite $\phi''(R)=-\alpha\phi(R)/R^2$, where
$\alpha>0$ is a proportionality constant. In the low energy
collision limit $E\rightarrow 0$, we get

\begin{equation}
V_{\mathrm{efm}}(R)=-\frac{\alpha\hbar^2}{mR^2}\,\,\,\,\,\,\mbox{for
$R<a$} \,, \label{efimov}
\end{equation}

A rigorous calculation performed by V. N. Efimov shows that with $R$
identified as the three-body hyperspherical radius, we have
$\alpha=s_0^2+\frac14$ and $s_0=1.00624...$ is a constant
\cite{Efimov1970}. For $R>|a|$, the effective potential is no longer
attractive \cite{Dincao2005}.

Right on two-body resonance $a\rightarrow\pm\infty$, the $-1/R^2$
Efimov potential extends to infinity and can support an infinite
number of three-body bound states (Efimov states); simple scaling
laws for the spatial extent $A_N$ and binding energy $E_N$ of the
$N$-th lowest Efimov state have been derived as

\begin{eqnarray}
A_N&=&\beta^N\times A^* \\
E_N&=&\beta^{-2N}\times E^* \, , \label{efimov}
\end{eqnarray}

\noindent where $\beta=e^{\pi/s_0}\approx22.7$ is a universal
constant \cite{Efimov1970}. These size and energy scaling laws are
among the most prominent universal features of Efimov's predictions.
Constants $A^*$ and $E^*$ depend on the three-body potential at
short range and are thus expected to be non-universal
\cite{Braaten2006}.

\subsection{Universality of Efimov physics near different Feshbach resonances}
\label{ef}

Recent observation of an Efimov resonance in the three-body
recombination process \cite{Esry1999} of ultracold cesium atoms
represents a major breakthrough in few-body physics
\cite{Kraemer2006}.

Here we suggest a new scheme to check the ``defacto'' universality
of Efimov physics implied by the expected slow variation of the
short-range three-body potential with magnetic field tuning. By
monitoring recombination loss near different, isolated
\textit{open-channel dominated} Feshbach resonances, we expect that
Efimov resonances of the same order can occur at the same scattering
lengths. Here we point out that the application of magnetic field
barely changes the interatomic potential in the entrance scattering
channel. We thus expect that, in the three-body sector, systems have
nearly identical off-resonant phase shift near different Feshbach
resonances.

To estimate the insensitivity of the open channel potential to
magnetic field, we note that the two-body background scattering
length varies less than 1 $a_0$ over 100~G at $a=2400$~$a_0$. (This
estimation is based on numerical calculation of cesium atom
scattering length in the highest triplet scattering channel, in
which Feshbach resonances do not exist.) This small variation can be
translated into a small fractional change of the scattering phase
shift by $|\delta\eta/\eta|<3\times 10^{-10}$ per Gauss
\cite{Chin2006}. This result suggests that the non-universal effects
of the three-body potential can potentially remain nearly unchanged
when the magnetic field is tuned to different Feshbach resonances.

\section{Universality in a dilute BEC with large scattering length:
Lee-Huang-Yang Corrections and beyond}

In a dilute Bose-Einstein condensate, the energy per particle is
given as $2\pi na\hbar^2/m$, which describes the fluid on length
scales longer than the coherence length $l=(16 na)^{-1/2}$. Due to
the weak coupling, corrections to the mean field term can be
calculated as expansions of a dimensionless parameter $a/l$, which
is in turn proportional to the diluteness parameter $\sqrt{na^3}$.
The energy per particle in a dilute homogeneous BEC is given by
\cite{Braaten2002}

\begin{equation}
\frac{E}{N}=\frac{2\pi \hbar^2
na}{m}[1+\frac{128}{15\sqrt{\pi}}\sqrt{na^3}+\frac{8(4\pi-3\sqrt{3})}
3na^3\ln na^3+Cna^3+...\,\,]\,, \label{lhy}
\end{equation}

\noindent where the lowest order contributions $\sqrt{na^3}$, called
the Lee-Huang-Yang (LHY) correction \cite{Lee1957}, and $na^3\ln
na^3$ term \cite{Wu1959} result from universal two- and three-body
correlations, and $C$ is a three-body parameter which depends on
three-body interactions and Efimov physics \cite{Braaten2002}.
Although Eq.~\ref{lhy} was originally derived based on a hard-sphere
potential, the validity of the LHY term for soft-sphere and
short-ranged attractive potentials has been numerically verified
\cite{Giorgini1999}.


Beyond mean-field effects can be amplified by tuning the scattering
length to large values. Previous approaches along this line with
$^{85}$Rb reached $na^3=0.1$, but were complicated by limited
lifetimes due to three-body inelastic collisions. Here we point out
that a careful choice of scattering length and a fast measurement
can allow for a detectable beyond mean-field signal.

To see this, we first note that the LHY term, on the order of
$(na^3)^{1/2}$, is a lower order process than is the three body
process of $na^3$. Measurement of the former effect can be immune
from three-body loss when $na^3$ is low. For example, the typical
mean-field energy of a BEC is $U=$h$\times$1~kHz and the scattering
length can be tuned such that $na^3=0.01$. In this case, the LHY
term is about $(na^3)^{1/2}=$ 10$\%$ of the mean-field energy and is
10 times larger than the three-body energy scale. The associated
three-body time scale is $na^3 U/\hbar\approx$ (10~ms)$^{-1}$.
Determination of interaction energy of a condensate within 10~ms can
be realized by promptly releasing the condensate into free space.
The expansion of the condensate thus converts the interaction energy
into detectable atomic kinetic energy.

\section{Experimental approach}

Two powerful experimental tools will be employed to explore few-body
physics: optical lattices to confine and isolate few atoms at each
lattice site in the Mott insulator phase, and magnetic Feshbach
resonances to control atomic interactions. Both can lead to precise
control of the few-body samples in different interaction regimes.

\subsection{Scattering properties of cesium atoms}
\begin{figure}
\begin{center}
\psfig{file=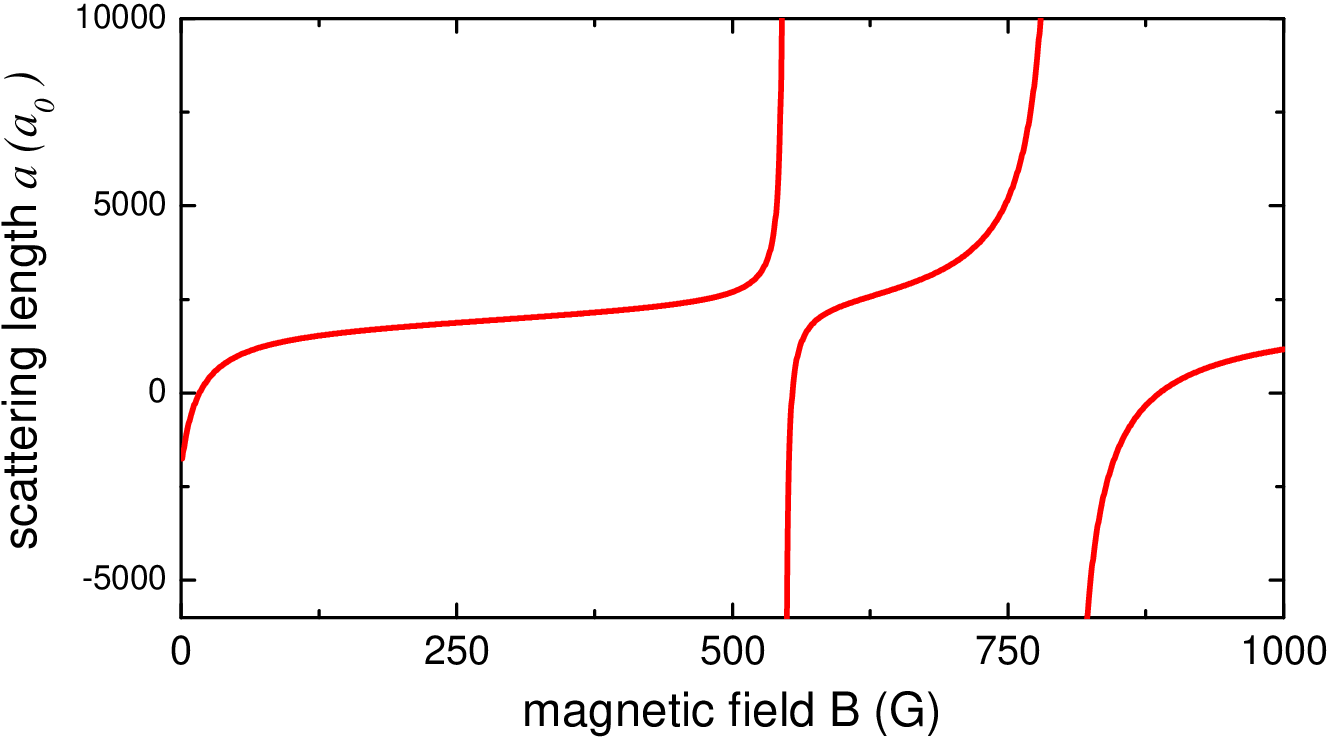,width=2.5in} \caption{Calculated $s$-wave
Feshbach resonances in collisions of ground state cesium atoms.
Three broad resonances at -11.7~G, 547~G and 800~G allow for tuning
of the scattering length. Other higher-order resonances are omitted
here for simplicity. The numerical calculation code is provided by
Eite Tiesinga, NIST.} \label{cs}
\end{center}
\end{figure}

Cesium-133 is chosen in the experiment for their convenient tuning
of interaction. In the range of 0 to 50~G, the s-wave scattering
length in the lowest hyperfine ground state $|F=3, m_F =3\rangle$
can be smoothly tuned from -2500~$a_0$ to 1000~$a_0$. Here, $F$ is
the total angular momentum quantum number and $m_F$ its projection
along the magnetic field. At higher fields, two more broad
resonances exist at 547~G and 800~G. See Fig.~\ref{cs}.

The existence of multiple broad $s$-wave Feshbach resonances permits
tests of universality by probing the cold atoms sample at different
scattering lengths. As discussed in Secs.~\ref{f} and \ref{ef},
unique tests of universality in two- and three-body systems can be
performed by tuning the scattering length to the same value, but
near different Feshbach resonances.

\subsection{Fast evaporation to Bose-Einstein condensation in optical traps}
We employ a novel scheme to achieve fast, runaway evaporative
cooling of cesium atoms in optical traps. This is realized by
tilting the optical potential with a magnetic field gradient.
Runaway evaporation is possible in this trap geometry due to the
very weak dependence of vibration frequencies on trap depth, which
preserves atomic density during the evaporation process. When the
trap depth is reduced by a large factor of 100, the geometric mean
of the trap frequencies is only reduced by a factor of 2 and thus
preserves the high collision rate \cite{Hung2008}.

Using this scheme, we show that Bose-Einstein condensation with
$\sim 10^5$ cesium atoms can be realized in $2\sim 4$~s of forced
evaporation \cite{Hung2008}. The evaporation speed and energetics
are consistent with the three-dimensional evaporation picture,
despite the fact that atoms can only leave the trap in the direction
of tilt.

\subsection{Preparation of few-atom systems in optical lattices}

Few-body experiments will begin with segmentation of a bulk
condensed superfluid into the ground states of isolated optical
lattice sites.  Each site will be populated with a small and in
general indefinite number, $1<N<10$, of atoms.

\begin{figure}
\begin{center}
\psfig{file=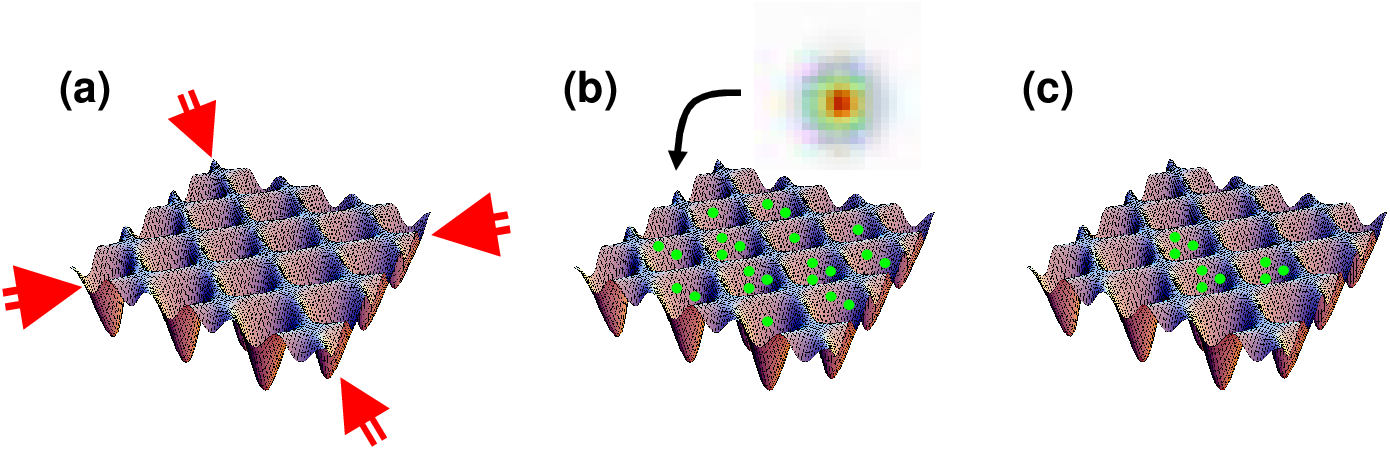,width=4in} \caption{Loading an optical
lattice and preparation of lattice sites with three atoms. (a)
Optical lattices are formed by the interference pattern of
intersecting laser beams. For a red-detuned lattice, the potential
minima are defined by the anti-nodes of the standing waves. (b)
Condensed atoms are loaded into the optical lattices. (c) To prepare
lattice sites with three and only three atoms, atoms in lattice
sites with other occupancies may be removed by precision
radio-frequency excitation.} \label{lattice}
\end{center}
\end{figure}

We have constructed a novel optical lattice configuration - a thin
layer, 2D optical lattice, which permits direct imaging of atomic
density by sending an imaging beam perpendicular to the lattice
plane. The optical lattice is defined by interfering four laser
beams derived from a single frequency fiber laser operated at a
wavelength $\lambda =1.06\mu$m. Two counterpropagating beam pairs on
the horizontal plane form a square optical lattice. In the vertical
direction, confinement is provided by a single CO$_2$ laser beam
focused to 50 $\mu$m vertically and 2~mm in the radial direction.
This tight vertical confinement holds atoms against gravity without
need for a magnetic field gradient, and provides an ideal
mode-matching potential for transferring condensates into the 2D
lattice, see Fig.~\ref{lattice}.

Probing of few-body energies will be performed through the combined
methods of precision radio-frequency spectroscopy, collective mode
excitation and dynamic evolution of matter wave coherence. In
particular, methods which allow precise determination of the
variation of few-body energies with atom number will permit direct
investigation of various interactions, including two- and three-body
scattering of free atoms, strong correlations, unitarity at strong
confinement, atom-dimer interactions, and three-body recombination.
Working in a tightly bound optical lattice allows quantum pressure
to determine atomic density profiles and permits accurate
extrapolation of single-particle measurements to interacting
few-body systems.  In addition, methods of adiabatically preparing
specific few-body systems as combinations of free and bound states
(e.g. atom+dimer) will be explored, providing a basis for directly
studying the universality of higher complexity interactions.

\section{Conclusion}

We describe key issues in three- and many-body physics including
Efimov physics and beyond mean-field effects in the context of
Feshbach tuning in quantum gases. In both cases, Bose-condensed
cesium atoms provide unique opportunities to investigate universal
behavior of energy shifts and energy structure. In particular, we
point out possible non-universal parameters, including the finite
interaction range $\bar{a}$, Feshbach coupling length scale $R^*$
and three-body phase shift.

We propose a brand new approach to prepare and study few-body
systems in optical lattices by inducing superfluid-Mott insulator
transition in a single layer of 2D optical lattices. This system
provides complete and independent control over the filling factor,
on site interaction and tunneling. Together with the rich
interaction properties of cesium atoms and fast evaporation, one
expect new level of few-body physics can be explored in this lattice
setting. We anticipate that a firm understanding of universality in
finite systems will provide practical applications in quantum
simulation of few-body systems in nuclear physics, helium physics,
physical chemistry and the physics of atom clusters.

\section{Acknowledgement}
The authors acknowledge support from the NSF-MRSEC program under No.
DMR-0213745, ARO Grant No. W911NF0710576 from the DARPA OLE program
and Packard foundation. N.G. acknowledges support from the Grainger
Foundation.


\begin{thebibliography}{1}
\bibitem{Inouye1998}
S. Inouye, M. R. Andrews, J. Stenger, H.-J. Miesner and D. M.
Stamper-Kurn and W. Ketterle, {\em Nature} {\bf 392}, 151 (1998).

\bibitem{Braaten2006}
E. Braaten and H.-W. Hammer,{\em Phys. Rep.} {\bf 428}, 259 (2006).

\bibitem{Regal2004}
C. A. Regal, M. Greiner and D. S. Jin, {\em Phys. Rev. Lett.} {\bf
92}, 040403 (2004).

\bibitem{Bartenstein2004}
M. Bartenstein, A. Altmeyer, S. Riedl, S. Jochim, C. Chin, J.
{Hecker Denschlag} and R. Grimm, {\em Phys. Rev. Lett.} {\bf 92},
120401 (2004).

\bibitem{Zwierlein2004}
M. W. Zwierlein, C. A. Stan, C. H. Schunck, S. M. F. Raupach, A. J.
Kerman and W. Ketterle, {\em Phys. Rev. Lett.} {\bf 92}, 120403
(2004).

\bibitem{Efimov1970}
V. Efimov, {\em Phys. Lett. B} {\bf 33}, 563 (1970).

\bibitem{Braaten2002}
E. Braaten, H.-W. Hammer, and T. S. Mehen, {\em Phys. Rev. Lett.}
{\bf 88}, 040401 (2002).

\bibitem{Tiesinga1993}
E. Tiesinga, B.J. Verhaar, B. J. and H.T.C. Stoof, {\em Phys. Rev.
A}, {\bf 47}, 4114 (1993).

\bibitem{Mott1965}
N. F. Mott and H. D. W. Massey, ``Theory of atomic collisions'',
{Oxford University Press, London}, (1965).

\bibitem{Julienne2006}
P. S. Julienne and B. Gao, {\em Atomic Physics 20}, Edt. by C. Roos
and H. H{\"a}ffner and R. Blatt, {AIP, Melville, New York}, pp.
261-268 (2006).

\bibitem{Gribakin1993}
G. F. Gribakin and V. V. Flambaum, {\em Phys. Rev. A} {\bf 48}, 546
(1993).

\bibitem{Chin2008}
C. Chin, R. Grimm, E. Tiesinga and P.S. Julienne, (to be submitted
to  {\em Rev. Mod. Phys.})

\bibitem{Petrov2004}
D. S. Petrov, Phys. Rev. Lett. {\bf 93}, 143201 (2004).

\bibitem{Dincao2005}
J. P. D'Incao and B. D. Esry, {\em Phys. Rev. Lett.} {\bf 94},
213201 (2005).

\bibitem{Esry1999}
B. D. Esry, C. H. Greene and J. P. Burke,{\em Phys. Rev. Lett.} {\bf
83}, 1751 (1999).

\bibitem{Kraemer2006}
T. Kraemer, M. Mark, P. Waldburger, J. G. Danzl, C. Chin, B.
Engeser, A.D. Lange, K. Pilch, A. Jaakkola, H.-C. N\"agerl and R.
Grimm, {\em Nature} {\bf 440}, 315 (2006).

\bibitem{Chin2006} C. Chin and V. V. Flambaum,
{\em Phys. Rev. Lett.} {\bf 96}, 230801 (2006).

\bibitem{Lee1957}
T. D. Lee, K. Huang, and C. N. Yang, {\em Phys. Rev.} {\bf 106},
1135 (1957).

\bibitem{Wu1959}
T. T. Wu, {\em Phys. Rev.} {\bf 115}, 1390 (1959).

\bibitem{Braaten2002}
E. Braaten, H.-W. Hammer, and T.S. Mehen, {\em Phys. Rev. Lett.}
{\bf 88}, 040401 (2002).

\bibitem{Giorgini1999}
S. Giorgini, J. Boronat, and J. Casulleras, {\em Phys. Rev. A} {\bf
60}, 5129 (1999).

\bibitem{Hung2008}
C.-L. Hung, X. Zhang, N. Gemelke, and C. Chin, {\em Phys. Rev. A}
{\bf 78}, 011604 (2008).


\end{thebibliography}
\end{document}